\documentclass[preprint,12pt]{revtex4}

 \usepackage{xcolor}
\usepackage{graphicx}
\usepackage{amssymb}
\usepackage{cancel}
\usepackage{ulem}
\usepackage{lineno}
\usepackage{amsmath}
\usepackage{appendix}

\begin{document}

\title{
 Random diffusivity scenarios behind \\ anomalous non-Gaussian  diffusion  
}

\author{M. A. F. dos Santos$^{1}$, E. H. Colombo$^{2,3}$ 
 and C. Anteneodo$^{1,4}$}
\address{$^{1}$ Department of Physics, PUC-Rio, Rua Marqu\^es de S\~ao Vicente 225, 22451-900, Rio de Janeiro, RJ, Brazil}
\address{$^{2}$ Department of Ecology \& Evolutionary Biology, \\ Princeton University, Princeton, NJ 08544, USA\looseness=-1}
\address{$^{3}$ Department of Ecology, Evolution, and Natural Resources, \\ Rutgers University,  New Brunswick, NJ 08901, USA\looseness=-1}
\address{$^{4}$ Institute of Science and Technology for Complex Systems, Brazil}

\begin{abstract}

The standard diffusive spreading, 
characterized by a Gaussian distribution with mean square displacement that grows linearly with time,  can   break down, for instance, under the presence of correlations and heterogeneity. 
In this work, we consider the spread of a population of fractional (long-time correlated) Brownian walkers,  with  time-dependent and heterogeneous diffusivity.
We aim to obtain the possible scenarios related to these individual-level features from the observation of the temporal evolution of the population spatial distribution. %
We develop and discuss the possibility and limitations of  this connection for the broad class of self-similar diffusion processes.
Our results are presented in terms of a general framework, which is then used to address  well-known processes, such as Laplace diffusion, nonlinear diffusion, and their extensions.
 \\

\end{abstract}

  

\maketitle

 
\section{Introduction} \label{sec:intro}

Diffusion processes are a fundamental ingredient of the dynamics of spatially-extended physical, chemical and biological systems. 
Regardless of the entity, from nanoparticles to complex biological organisms, it is common that, due to the lack of coherence in external drivers, random, uncorrelated  motion can arise. 
In the case of Brownian motion (Bm), 
the spreading of a given substance follows normal diffusion, characterized by (i) a Gaussian distribution, (ii) with mean square displacement (MSD) that grows linearly with time. 

These two remarkable properties are quite ubiquitous in nature, but tend to break down under more complex scenarios. 
For instance, a class of situations produce the so-called Brownian yet non-Gaussian behavior, which implies a deviation from the Gaussian shape, but preserving the linear law of the MSD~\cite{wang2009anomalous}. In contrast, in other cases, the Gaussian shape can be preserved, while the mean square displacement, $\langle {\bf r}^2 \rangle $, displays an anomalous  growth with time, $t$, increasing as a power-law with exponent different from one, i.e., $\langle {\bf r}^2 \rangle \propto t^{\gamma}$ with $\gamma\neq 1$, as occurs in the cases of   fractional Brownian motion (fBm)~\cite{mandelbrot1968fractional} or scaled diffusion~\cite{lim2002self}. 

FBm emerges when long-time correlations are present in the microscopic forces. Therefore, in isotropic media, the  $d$-dimensional position vector ${\bf r}=(x_1,\ldots,x_d)$  of a tracer  is driven by $\dot{\bf r}(t)= \sqrt{2D } \, {\pmb \eta}_H(t)$, where $D$ is the  diffusivity and ${\pmb \eta}_H(t)$ is a vector whose components are  fractional Gaussian noises with zero mean and  correlations $\langle {  \eta}_i(t){  \eta}_j(t+\tau)\rangle \sim \delta_{ij}\tau^{2(H-1)}$, for $\tau\neq 0$,  characterized by the Hurst exponent $H$~\cite{mandelbrot1968fractional}. As a consequence, the probability distribution associated to  tracers starting at the origin is given by
\begin{eqnarray}
\mathcal{G}_{H}({\bf r},t|D) = \frac{1}{(4 \pi D\, t^{2H})^{\frac{d}{2}}}  \exp\Bigl( - \displaystyle \frac{{\bf r}^2}{4 D \,t^{2H}} \Bigr)\,, \label{eq:fBMgaussian}
\end{eqnarray}
where $d$ is the spatial dimension. 
That is,    the shape is Gaussian in the spatial coordinates $x_i$, with $i=1,\ldots,d$,  
which scale with time as $t^H$. For, $H=1/2$, ${\pmb \eta}_H(t)$ is the vector of white noise components, thus yielding normal diffusion. But, in general, for $2H\neq 1$, the diffusive spreading is Gaussian but anomalous. 
Experimentally, the behavior described by  Eq.~(\ref{eq:fBMgaussian}) has been observed for diverse  values of $H$, for instance, for the diffusion of nanometric beads in crowded environments, lipid granules and telomeres, among  several other cases~\cite{ernst2012fractional,jeon2011vivo,PhysRevE.83.041919,burnecki2012universal}.  

However, the picture provided by Eq.~(\ref{eq:fBMgaussian}) can change, for instance, if $D$ depends on time and  sources of heterogeneity are present, 
which can have a substantial impact on the statistical properties of the ensemble. 
A time-dependent diffusion coefficient can arise due  to intrinsic or extrinsic factors, related to organisms memory and aging effects or due to transformation in the environment spatial structure~\cite{latour1994time,sen2004time,safdari2015aging,novikov2019quantifying,cherstvy2021anomalous}.
These effects can be incorporated through scaled diffusion~\cite{lim2002self} by modulating the diffusivity with a function  $\lambda(t)$.

Besides time-dependency, heterogeneity is also  quite ubiquitous in real systems. Its impact can be addressed  through the superstatistical approach proposed by C. Beck and E. G. D. Cohen to extend statistical mechanics to complex heterogeneous environments. 
Superstatistics has been applied to run-and-tumble particles~\cite{sevilla2019stationary}, animal movement~\cite{petrovskii2009dispersal,hapca2009anomalous,santana2020langevin},  metapopulation extinction dynamics~\cite{colombo2019connecting}, time series  analyses~\cite{beck2005time,mendes2007statistics,cortines2008measurable,anteneodo2009statistical}, and many other cases~\cite{garcia2011superstatistics,hidalgo2021cusp,dos2021probability,agahi2020truncated,dos2020mittag,dos2020log,rominger2019nonequilibrium}.
The superstatistics of fBm has been recently developed, 
providing theoretical support for various experimental observations, such as, protein diffusion in bacteria \cite{beck2021superstatistical,sadoon2018anomalous}, micro-particles in a bi-dimensional system with disordered distribution of pillars \cite{chakraborty2020disorder}, and tracer diffusion in mucin hydrogels \cite{cherstvy2019non} (see Refs.~\cite{sabri2020elucidating,gires2020quantifying,jeon2016protein,cherstvy2018non,lampo2017cytoplasmic} for more examples).  

Throughout the next sections, we incorporate these different sources  of deviation from the Bm picture  and study the resulting   overall probability density function (PDF), 
\begin{eqnarray}
p({\bf r},t) &=& \int_0^{\infty} dD \;\pi(D,t) \,\mathcal{G}_{H}({\bf r},t|D),  \label{eq:superstatisticalScaled}
\end{eqnarray}
which describes the spatial distribution of the tracers,  where  $\mathcal{G}_{H}({\bf r},t|D)$ was defined in Eq. (\ref{eq:fBMgaussian}) and 
$\pi(D,t)$ is the PDF of the diffusivity $D$, with parameters that are function of time ~\cite{wang2021ergodic}.   A similar superstatistical approach was investigated in  connection with the {\it generalized gray Brownian process}, from the point of view of  ergodicity breaking~ \cite{molina2016fractional}.

Our aim in the following sections is to obtain the reverse path, accessing the 
microscopic variability expressed by $\pi(D,t)$,  from the macroscopic observable  $p({\bf r},t)$. While the shape of the distribution (e.g., exponential, Gaussian, power-law, etc.) of diffusivities, $\pi(D,t)$, can be identified, 
how it spreads depends on the interplay between 
scaled and fractional diffusion which become blended by the superstatistical procedure, as we will see soon.

Furthermore, we explore the fact that we  can reconstruct the macroscopic properties of other classes of diffusion processes, such as those embodied in the nonlinear diffusion equation, known to be generated by density-dependent feedbacks~\cite{colombo2018}. Thus, our results highlight the impact of heterogeneity and provide warnings for attempts to experimentally infer the microscopic dynamics of a given system. For particular cases, we perform numerical simulations of large heterogeneous populations, 
to illustrate and contextualize our analytical results.

The  paper is organized as follows.
In Sec.~\ref{sec:approach}, we present a procedure to obtain the microscopic variability from the macroscopic level, and we also provide two simple examples to illustrate the procedure. 
In Sec.~\ref{sec:generalized}, we apply the theory to two general families of macroscopic distributions: stretched exponential and power-law ones. 
In Sec.~\ref{sec:final}, we present remarks and conclusions.

\section{From the microscopic to the macroscopic level and the other way around}
\label{sec:approach}

In this section, we first perform the superstatistical mixture, as defined by  Eq.~(\ref{eq:superstatisticalScaled}), obtaining the overall PDF  $p(\mathbf{r},t)$, for known  microscopic variability and time-dependence of the diffusion coefficient, expressed in $\pi(D,t)$. 
Then, we show how to revert this process, hence allowing to extract  $\pi(D,t)$ from $p(\mathbf{r},t)$.
We consider the broad scenario of self-similar diffusion processes for which $p(\mathbf{r},t)$ is a function of the  scaled variable $\xi \equiv  |{\bf r}|/t^{\gamma/2}$,
\begin{eqnarray}
p({\bf r},t) = \frac{1}{t^{\gamma d/2}} F\left( |{\bf r}|/t^{\gamma/2}  \right), 
\label{eq:scalingdiffusionlaw}
\end{eqnarray}
where the scaling function $F$ has an arbitrary but normalizable form.
Some examples are: L\'evy processes \cite{uchaikin2003self}, nonlinear diffusion \cite{bologna2000anomalous}, fractional diffusion \cite{bologna2000anomalous,pagnini2012erdelyi}, among others \cite{lim2002self,uchaikin2002subordinated,uchaikin2003self,ferrari2001strongly,oliveira2019anomalous,Santos2019analytic,mainardi2005fox} that produce non-Gaussian (lepto and platykurtic)  forms. 

In the current scenario, we assume that $p({\bf r},t)$ is generated by a collection of fB walkers, i.e., ruled by Eq.~(\ref{eq:fBMgaussian}), with distinct diffusion coefficients which can change with time, $D_i = \mu_i \lambda(t)$, where $\mu_i$, is a mobility constant and $\lambda(t)$ is, in principle, an arbitrary deterministic function that is the same for all individuals. 
The values of $\{\mu_i\}$ are assumed to be sampled from an unknown PDF, $\theta(\mu)$. As a consequence,  in the large population limit, $N\gg 1$, diffusivity variability is well described by the PDF, $\pi(D,t) = \theta(\mu)/\lambda(t)$.

The aim of the following calculation is to uncover $\pi(D,t)$ from the ensemble PDF, $p(\mathbf{r},t)$. To reach this result, we first apply the superstatistics recipe 
and then invert the procedure. As we will see, the connection between PDFs will be done through Laplace transforms, thus, it is convenient (without loss of generality) to express $\pi(D,t)$  as
\begin{eqnarray}
\pi(D,t) \equiv \frac{1}{\lambda(t)}\pi_s\left(\frac{\lambda(t)}{D}\right), \label{eq:scaled-diffusivity}
\end{eqnarray}
where $\pi_s(y)$ is an  auxiliary function that depends only on  the  scaled variable $y\equiv \lambda(t)/D = 1/\mu$. 
Substituting 
Eq.~(\ref{eq:scaled-diffusivity}) 
and Eq.~(\ref{eq:fBMgaussian}) into Eq.~(\ref{eq:superstatisticalScaled}), we obtain  
\begin{equation}
p({\bf r},t) =  \int_{0}^{\infty} \frac{\Bar{\pi}(y)}{ ( 4 \lambda(t)\,  t^{2H})^{\frac{d}{2}} }\exp\Bigl( - \displaystyle  y \frac{{\bf r}^2}{ 4\lambda(t)\, t^{2H}} \Bigr) dy\, , 
\label{eq:pdfdistribution1}
\end{equation}
with
\begin{eqnarray}
\Bar{\pi}(y)=  \frac{\pi_{s}\left(y \right)}{\pi^{\frac{d}{2}}y^{2-\frac{d}{2}} }. \label{eq:upsilon}
\end{eqnarray}
It is central to notice that the integral over $y$  in Eq.~(\ref{eq:pdfdistribution1}) has exactly the Laplace transform structure, i.e.,  it has the form $\mathcal{L}_{y\to s}\left\{\Bar{\pi}(y)\right\} = \int_0^{\infty} e^{-s y}\Bar{\pi}(y) dy $,  
which maps $y$ to a  new variable $s= \xi^2={\bf r}^2/[4\lambda(t)\,t^{2H}]$.  

Although there are many classes of scaled diffusion, with general time scaling $\lambda(t)$~\cite{sposini2018random,dos2021random},  self-similarity demands  for a specific choice of $\lambda(t)$, to recover the scaling form of Eq.~(\ref{eq:scalingdiffusionlaw}), then we restrict the form of  $\lambda(t)$   to a power-law function of time, namely, 
\begin{eqnarray}
4\lambda(t) =   t^{\alpha-1}.
\label{eq:lambda}
\end{eqnarray}
This form embraces numerous experiments across disciplines that include the impact of aging or parameter time-dependency. For instance,   physical properties of the medium, such as viscosity and temperature, can directly influence the diffusion processes and can change over time under non-stationary constraints~\cite{cherstvy2021anomalous}. Also, time-dependency can be a probe to medium topology~\cite{sen2004time}. In more general settings, biological and chemical factors can be at play and influence diffusion in a complex manner~(see Ref. \cite{munoz2021objective} for a review).

Under the choice of $\lambda(t)$ set in Eq.~(\ref{eq:lambda}), Eq.~(\ref{eq:pdfdistribution1}) becomes
\begin{equation}
p({\bf r},t) =  \frac{1}{t^{(2H+\alpha-1)d/2}}
\underbrace{\int_{0}^{\infty}  \Bar{\pi}(y) \exp\Bigl( - \displaystyle  y \frac{{\bf r}^2}{  t^{2H+\alpha-1}} \Bigr) dy}_{\mathcal{L}  \left\{\Bar{\pi}(y) \right\}}\, , 
\label{eq:pdfdistribution2}
\end{equation}
which, compared to Eq.~(\ref{eq:scalingdiffusionlaw}), allows to identify 
\begin{equation} \label{eq:gamma}
\gamma  = 2H+\alpha-1\,,
\end{equation}
and
\begin{equation} \label{eq:F}
F(\xi) = \mathcal{L}_{ y \to \xi^2  } \left\{\Bar{\pi}(y) \right\}  \,,
\end{equation}
with $\xi=|{\bf r}|/t^{\gamma/2}$. 
That is, we can  rewrite Eq. (\ref{eq:pdfdistribution1}) in compact form as
\begin{eqnarray}
p({\bf r},t) = \frac{1}{t^{\gamma d/2} } \mathcal{L}_{ y \to \xi^2 } \left\{\Bar{\pi}(y) \right\} \Big|_{ \displaystyle \xi = \frac{| {\bf r}|}{t^{\gamma/2  }} }.
\label{eq:generalsolution}
\end{eqnarray}

First, note, from Eq.~(\ref{eq:superstatisticalScaled}), that, if all the walkers  have the same mobility, $\Bar{D}(t)$, that is, $\pi(D,t) = \delta(D-\Bar{D}(t))$, the population distribution has the Gaussian shape. 
Also, note that, if $\gamma \equiv 2H+\alpha-1 = 1$, then the MSD grows linearly with time.

In fact, the  $\mu$th  moment $\langle |{\bf r}|^{\mu} \rangle$ evolves according to $
\langle |{\bf r}|^{\mu} \rangle \approx \mathcal{C}_{\mu,d} \, t^{(2H+\alpha-1) \mu/2}
$ (see \ref{app:anomalous}), thus the growth of the second moment (i.e., $\mu = 2$) follows $\langle |{\bf r}|^2 \rangle \sim t^{2H+\alpha-1}$. 
Therefore, anomalous diffusion can be caused by correlations ruled by the exponent $H$, and/or time-dependency, via the exponent $\alpha$. 
For instance, the effect of anticorrelated motion ($H<0.5$) can be compensated by a sufficiently fast increase with time ($\alpha>1$).

Equation~(\ref{eq:generalsolution}) is the bottom-up connection  providing the macroscopic-level patterns from the microscopic-level dynamics. Then, exploring Laplace-transform identifications, we can obtain a top-down connection, obtaining the micro from the macro levels, following the steps:  
(i) we write the overall PDF  $p({\bf r},t)$ in the self-similar form (\ref{eq:scalingdiffusionlaw}),  such that it  will allow us to find  the function $F(\xi)=\mathcal{L}_{y\to \xi^2}\{\Bar{\pi}(y) \}$;  
(ii)  by means of   Eq.~(\ref{eq:generalsolution}), we perform the inverse Laplace transform $\mathcal{L}^{-1}_{s\to y}\{F(\sqrt{s})\}$ with $s=\xi^2$, to  identify the $\overline{\pi}(y)$, and  
by using Eq.~(\ref{eq:upsilon}), we get $\pi_s(y)$;
(iii) finally we arrive at $\pi(D,t)$ by means of  Eq.~(\ref{eq:scaled-diffusivity})). 
This entire procedure is summarized by the following expression
\begin{eqnarray}
\pi(D,t) =\frac{( 4\pi t^{2H} )^{\frac{d}{2}} t^{\alpha-1} }{4 D^{2-\frac{d}{2}}}\mathcal{L}^{-1}_{s \to y} \left\{ p(\xi=\sqrt{s},t) \right\}  \Bigg|_{\displaystyle y= \frac{t^{\alpha-1}}{4D} }, \label{eq:Superdiffusivity}
\end{eqnarray}
recalling that $\xi=|{\bf r}|/t^{(2H+\alpha-1)/2}$ and the inverse Laplace operator leads us to the change $s \xrightarrow{\mathcal{L}^{-1}} y$.  For different scaling functions $F(\xi)$ in Eq. (\ref{eq:scalingdiffusionlaw}), different forms of $\pi(D,t)$    can emerge, such as,   exponential  \cite{wang2012brownian,wang2009anomalous},   $\chi^2$-Gamma \cite{sposini2018random,dos2021random} and Rayleigh PDFs~\cite{cherstvy2016anomalous}.

Importantly, we are assuming that diffusivities differ between individuals, but change deterministically over time with the same scaling factor. If the diffusivity were stochastic (producing the same $ \pi(D, t) $), then, the ensemble distribution, $ p $, would necessarily have a crossover to Gaussianity at long times, exhibiting an ensemble  diffusion coefficient which is simply the average  $D = N^{-1}\sum_i^ND_i$. 
However, for sufficiently short observation times, the superstatistical picture is present and, as a consequence, our results could be applied to this regime (for further discussion, see Refs.~\cite {chechkin2017brownian, wang2009anomalous}).

In the following subsections, we provide two concrete simple examples in 1D, namely, related to Laplace  and nonlinear diffusion, respectively.

\subsection{Laplace diffusion} 
\label{sec:laplace}

Laplace diffusion represents a process with an 
exponential spread in space,   which in 1D is 
characterized by 
$p(x,t) \propto \exp(-|x|/\lambda(t))$, 
being $\lambda(t)$ a time-dependent scaling~\cite{metzler2020superstatistics,postnikov2020brownian}. It has been reported for  tracers diffusing in complex fluids,  
such as 
%
particles with fluctuating size~\cite{hidalgo2020hitchhiker} or hard-sphere colloidal suspensions~\cite{guan2014even},  in the anomalous spreading dynamics of amoeboid cells~\cite{cherstvy2018non}, protein crowding in lipid bilayers~\cite{jeon2016protein}, and in  diffusing diffusivity  models~\cite{chechkin2017brownian,chubynsky2014diffusing}.  
There are also experiments where Laplace diffusion emerges, although  only during timescales shorter than the correlation times of tracers~\cite{wang2009anomalous,wang2012brownian}.  

For our present purpose, let us consider a Laplace diffusion with power-law temporal scaling, given by
\begin{eqnarray}
p(x,t)  = \frac{1}{\sqrt{4 D_0 \, t^{\gamma}}} \exp \Bigl( -  \frac{|x|}{\sqrt{D_0 \, t^{\gamma}}} \Bigr), \label{eq:distributionLaplace}
\end{eqnarray}
in which $D_0$ is the ensemble diffusivity and $\gamma$ a parameter that characterizes the dynamics, such that 
the second moment is given by $\langle x^2 \rangle = 2 D_0 t^{\gamma}$. 
To uncover the superstatistics behind Eq.~(\ref{eq:distributionLaplace}), 
we rewrite it in a self-similar form, in terms of the invariant variable $\xi=|x|/t^{\frac{\gamma}{2}}$, namely,
\begin{eqnarray}
p(\xi,t) =  \frac{1}{\sqrt{4 D_0 t^{\gamma}} } \exp \Bigl( - \frac{\xi}{\sqrt{D_0}} \Bigr) ,
\label{eq:pxi1}
\end{eqnarray}
which from Eq.~(\ref{eq:gamma}) allows us to identify  $\gamma = 2H+\alpha-1$.  
By substituting  
Eq.~(\ref{eq:pxi1}) into Eq.~(\ref{eq:Superdiffusivity}) with $d=1$, 
we find the diffusivity distribution  
\begin{eqnarray}
\pi(D,t) & = & \frac{\sqrt{\pi t^{\alpha-1}} }{4 D^{\frac{3}{2}}\sqrt{D_0 }}\mathcal{L}^{-1}_{s \to y} \left\{ \exp \Bigl( - \sqrt{\frac{s}{D_0}} \Bigr) \right\}  \Bigg|_{\displaystyle y= \frac{t^{\alpha-1}}{4D} }  \nonumber \\ 
& = &  \frac{t^{1-\alpha}}{D_0} \exp\Bigl( -  t^{1-\alpha}\frac{D}{D_0} \Bigr),\label{eq:diffusivityLaplace}
\end{eqnarray}
recalling that the inversion of the Laplace transform needs to be performed on the variable $s=\xi^2$. 
This simple example illustrates the treatment applied to the overall PDF  of tracers to obtain the PDF of diffusivities. 
It is worth noting that exponential PDF of diffusivities, with time dependence, has been previously considered to analyze the ergodicity of tracer diffusion with uncorrelated noise~\cite{wang2021ergodic}.

\begin{figure}[h!]  
	\centering
	\includegraphics[width=1.0\textwidth]{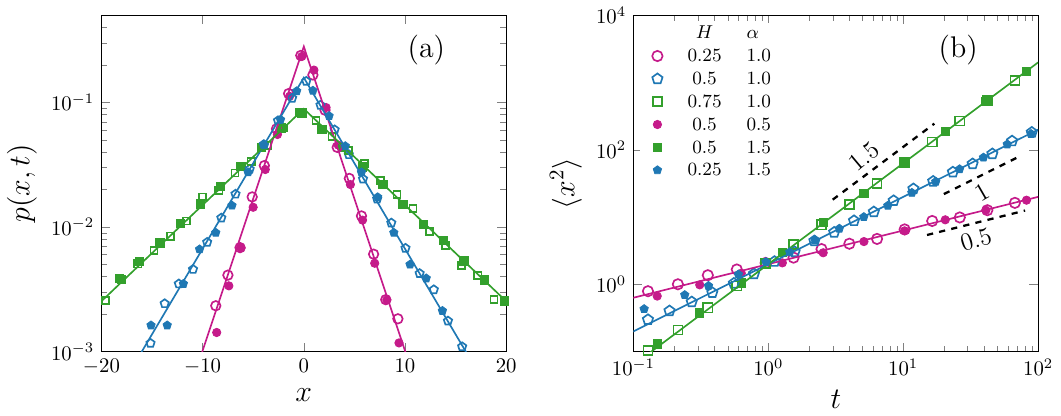}
	\caption{ 
		{\bf Laplace diffusion.} a) Probability density function of tracers $p(x,t)$ vs. position $x$, at time $t=10$. 
		(b) Mean square displacement $\langle x^2 \rangle$ vs. time $t$. 
		Solid lines are given by 
		Eq.~(\ref{eq:distributionLaplace}), with 
		$D_0=1$, and different values of $\gamma=2H+\alpha-1$. 
		Symbols are the result of numerical simulations ($2 \;10^4$ trajectories) using the corresponding values of the Hurst exponent, $H$, and the time scaling exponent, $\alpha$ (see~\ref{app:num}).
	} 
	\label{fig:laplace}
\end{figure}

To exercise our theoretical procedure, let us see  how  to  obtain  the  diffusivity  distribution  from  virtual  experiments. 
We assume that single-particle tracking is not available, thus, population spreading is characterized by a  population-level picture as depicted in Fig.~\ref{fig:laplace}. 
The overall PDF at time instant $t=10$  and the temporal evolution of the MSD, associated to Laplace diffusion, are shown  for different values of $H$ and $\alpha$. 
These parameters and diffusivity variability are in principle unknown, since we 
assume  that individual-level information is not accessible. 
However, our derivations using inverse Laplace transform show that it is possible to reveal the PDF of diffusivities $\pi(D,t)$ from the shape and scaling of the MSD. 
In this Laplace diffusion case, the distribution is exponential 
but notice that, pairs ($H$,$\alpha$) giving the same value of $\gamma$  yield identical population-level results, 
creating ambiguity about  the microscopic dynamics.

\begin{figure}[h]  
	\centering
	\includegraphics[width=1.0\textwidth]{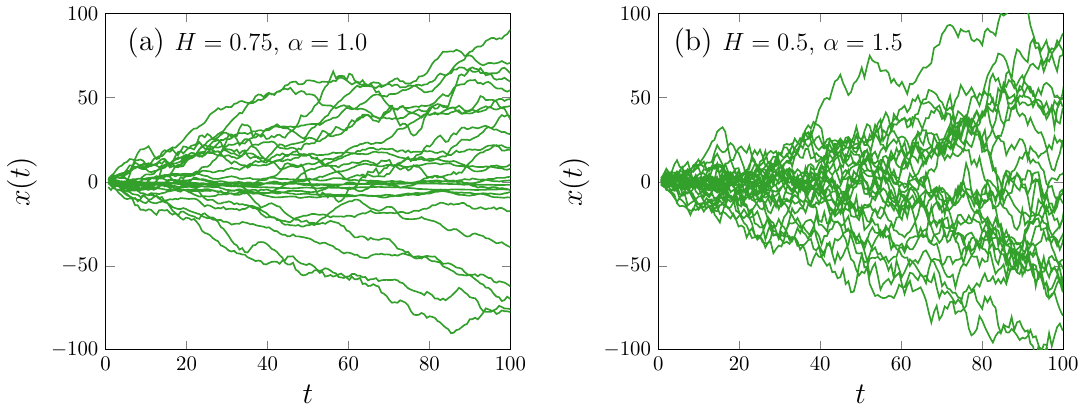}
	\caption{ 
		Fractional Brownian walks with scaled diffusion, with the values of $H$ and $\alpha$ indicated in the legends, such that $\gamma=2H + \alpha - 1 = 1.5$. Twenty trajectories are plotted in each case. 
		The diffusivities $D$ were sorted according to the exponential distribution  in Eq. (\ref{eq:diffusivityLaplace}) with $D_0=1$. 
		Although the trajectories have visibly different nature, they lead to the same $p(x,t)$ given by Eq.~(\ref{eq:distributionLaplace}), which characterizes Laplace diffusion. 
	}
	\label{fig:trajectories}
\end{figure}

In order to explore the range of microscopic dynamics 
constrained by the relation $\gamma= 2H + \alpha -1$, 
we perform numerical simulations of a collection of fB walker. 
For instance, let us consider the cases in which $\gamma = 1.5$ in Fig.~\ref{fig:laplace}, namely, $(H, \alpha) =  (0.75,1.0)$  and $(H, \alpha) =  (0.5,1.5)$.  
For both cases, a representative sample of trajectories is shown in Fig.~\ref{fig:trajectories}. 
Each trajectory is independent and governed 
by a suitable Langevin equation, which accounts for time-dependent diffusivity and fractional noise (see~\ref{app:num}).   
The diffusivity of each particle is chosen from the found exponential distribution Eq.~(\ref{eq:diffusivityLaplace}), with $D_0=1$ in all cases.

Note that despite the fact that the values of $H$ are different (also visible in the roughness of the trajectories),  diffusion time-dependency compensates its effects, resulting  that, in both cases, the MSD grows with the same exponent $\gamma$.
Actually, note that this ambiguity comes from the particular choice of a power-law $\lambda(t)$, needed to ensure self-similarity.  
Examples of underlying scenarios with static ($\alpha=1$, empty symbols) and time-dependent ($\alpha\neq 1$, filled symbols) diffusivity are given.

\subsection{Nonlinear diffusion}
\label{sec:porous}

The nonlinear diffusion equation in 1D is defined by 
\begin{eqnarray}
\frac{\partial \ }{\partial t} p(x,t) = D_0\frac{\partial^2 \ }{\partial x^2 } \left\{ p(x,t) \right\}^{m}, \label{eq:porousmedium}
\end{eqnarray}
where $ D_0 $ is the  diffusivity coefficient, and $ m>-1$ is a phenomenological exponent. 
For $m=1$, Eq.~(\ref{eq:porousmedium}) leads to the standard diffusion equation associated to Bm. 
For $m\neq 1$, it phenomenologically describes the dispersion
with a density-dependent diffusion coefficient that can be associated to 
internal feedbacks~\cite{dornelas2019single}.

For $m>1$, Eq.~(\ref{eq:porousmedium}) is known as porous media equation~\cite{peletier1981porous} which implies Barenblatt-Pattle solutions, and it is associated with  subdiffusion and a compact-support PDF~\cite{muskat1938flow,polubarinova2015theory}.
Eq.~(\ref{eq:porousmedium}) can be  extended to the full interval of $m$ in which the solution is normalizable, that is, for $m>-1$,  what  has been theoretically considered in diverse scenarios of statistical physics~\cite{frank2005nonlinear,abe2007superstatistics,
	plastino1995non,tsallis1996anomalous,mendes2015nonlinear,anteneodo2005non,schwammle2008q,Metzler2021,santos2021microscopic,gravanis2021stochastic,lavrova2021barenblatt}.
Cases well described by $m<1$  have been   experimentally observed, in confined granular systems~\cite{combe2015experimental} and in biological populations~\cite{mendes2015nonlinear}.
Since, we consider mixtures of Gaussian shapes (with tails that extend indefinitely), the resulting PDF of the population cannot become a  compact support shape (as those yield by $m>1$). Thus, we address only the interval   $m\in(-1,1)$, which is associated to distributions with a power-law tail, having infinite   ($-1<m<1/3$) or finite  variance ($1/3<m<1$), and always superdiffusive spreading 
$x^2\sim t^{2/(1+m)}$.

Restricting  $m \in(-1,1]$, and
using the change of exponents $m=1-1/\nu$, 
hence $\nu \in (1/2,\infty)$, the solution
of Eq. (\ref{eq:porousmedium}) can be written as 
\begin{eqnarray}
p(x,t) & = & \frac{1}{\gamma_{\nu}\beta_{\nu} t^{\frac{\nu}{2\nu-1}} }
\frac{1}{\left( 1+ \frac{1}{\nu\beta^2_{\nu}} \frac{x^2}{ t^{\frac{2\nu}{2\nu-1}}} \right)^{\nu}}.   \label{eq:nonlinearDiffusion}
\end{eqnarray}
This normalized shape is known as $q$-Gaussian within the frame of nonextensive  Tsallis statistics~\cite{tsallis2009introduction}.

Importantly,  notice that, at the same time that the exponent $\nu$ controls non-Gaussianity, it rules anomalous diffusion.  The superdiffusive spreading is evident from the scaling   $x^2\sim t^{2\nu/(2\nu-1)}$ in Eq.~(\ref{eq:nonlinearDiffusion}), 
since $2\nu/(2\nu-1)>1$ for $\nu>1/2$.
Notice that this scaling holds even when the MSD is divergent (i.e., for $1/2<\nu<3/2$). 

The remaining coefficients in Eq.~(\ref{eq:nonlinearDiffusion}) are defined by $\beta_{\nu}=[2\nu^{-2}(\nu-1)(2\nu-1)\gamma_{\nu}^{\frac{1}{\nu}}D_0]^{\frac{\nu}{2\nu-1}}$ and $\gamma_{\nu}= \sqrt{\nu\pi}\, \Gamma\left( \frac{2\nu-1}{2}\right)/\Gamma(\nu)$. 
The Gaussian shape is recovered in the limit $\nu\to \infty $, while the limit $\nu \to 1$ leads to the Cauchy-Lorentz distribution. 
Particularly, the Cauchy-Lorentz solution 
also results from the diffusion equation with $\mu$-fractional Laplacian, taking $\mu=1$~\cite{metzler2000random}. 

To show that the PDF of tracers can emerge from a mixing of fractional walkers on a disordered medium, we start by writing the  solution (\ref{eq:nonlinearDiffusion}) in a  self-similar form, by considering  $\xi=|x|/t^{\frac{\nu}{2\nu-1}}$ in Eq.~(\ref{eq:nonlinearDiffusion}). 
After comparison with Eq. (\ref{eq:generalsolution}), we obtain
\begin{eqnarray}
p(\xi,t) & = & \frac{1}{\gamma_{\nu}\beta_{\nu} t^{(2H+\alpha-1)/2}}
\frac{1}{\left( 1+ \frac{\xi^2}{\nu\beta^2_{\nu}}  \right)^{\nu}}\,, \label{eq:nonlinearDiffusion1}
\end{eqnarray}
where we have identified $2H+\alpha-1=\frac{2\nu}{2\nu-1}$.  
In this case,   $H$ and $\alpha$  are tangled through   the exponent $\nu$  that controls 
both anomalous diffusion and  non-Gaussianity, 
differently to the previous example, 
where they affected only the exponent of diffusive spreading $\gamma$.

To find the statistics of diffusivities, we substitute  Eq.~(\ref{eq:nonlinearDiffusion1}) into Eq.~(\ref{eq:Superdiffusivity}), leading to
\begin{eqnarray}
\pi(D,t) & = &
\frac{( 4\pi t^{\alpha-1} )^{\frac{1}{2}}  }{4 D^{\frac{3}{2}}\gamma_{\nu}\beta_{\nu}}\mathcal{L}^{-1}_{s\to y} \left\{ 1/\Bigl( 1+ \frac{s}{\nu\beta^2_{\nu}}  \Bigr)^{\nu} \right\}  \Bigg|_{\displaystyle y= \frac{t^{\alpha-1}}{4D} },
\label{eq:piinverseNon} 
\end{eqnarray}
with $s=\xi^2$. Performing the inverse Laplace transform in Eq.~(\ref{eq:piinverseNon}), we arrive at
\begin{eqnarray}
\pi(D,t) & = &  \frac{1}{ \Gamma(\overline{\nu}) } \frac{\left(\nu t^{(\alpha-1)}\beta_{\nu}^2/4  \right)^{\overline{\nu}} }{D^{1+\overline{\nu}}} \exp\Bigl(-\frac{\nu t^{(\alpha-1)} \beta_{\nu}^2 }{4D}\Bigr) ,  \label{eq:diffusivityPorous}
\end{eqnarray}
where $\overline{\nu}=\nu-1/2=1/(2[2H+\alpha-2])$ (see~\ref{app:porous}). This result reveals that  the superstatistics behind nonlinear diffusion gives an inverse $\chi^2$-Gamma distribution. Likewise in the previous section, we highlight two distinct cases: 
(i)  for $\alpha=1$ and hence $\gamma =2H=2\nu/(2\nu -1)$, we have fractional tracers moving in a medium described by a static PDF of diffusivities; (ii) for $H=1/2$, hence  $\gamma=\alpha=2\nu/(2\nu -1)$, we have  Brownian (non correlated) tracers moving on a medium that changes over time. Particularly, let us note that the former situation with $H=\nu/(2\nu-1)$ was previously reported~\cite{borland1998microscopic}, although with a non-Gaussian shape generated by a stochastic process with position-dependent diffusion coefficient. 

\begin{figure}[h!]  
	\centering
	\includegraphics[width=1.0\textwidth]{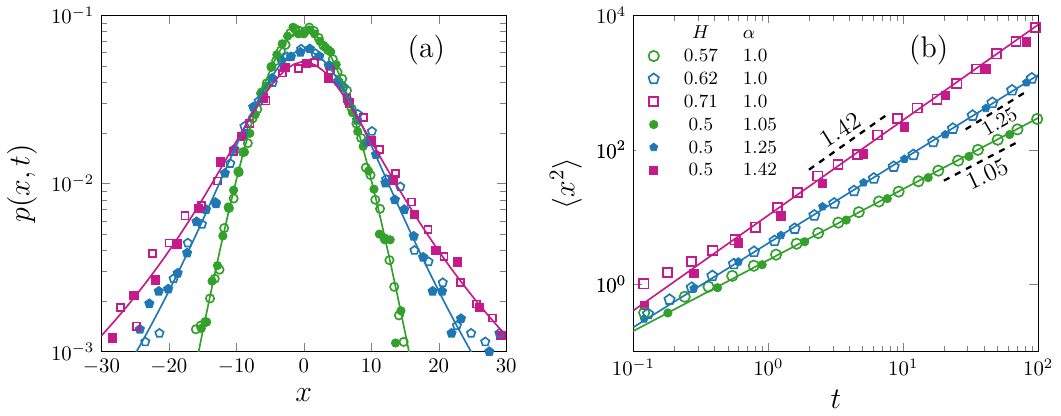}
	\caption{ 
		{\bf Nonlinear diffusion.} 
		(a) Probability density function of tracers $p(x,t)$ vs. position $x$, at time $t=10$. 
		(b) Mean square displacement $\langle x^2 \rangle$ vs. time $t$. 
		In both cases, we considered $D_0=1$ and heterogeneous tracers. Different values of exponents $\nu$  (recalling that $H=\nu/(2\nu-1)$ and
		and $\alpha$ were considered.
		Symbols are the result of numerical simulations ($2 \;10^4$ trajectories). Solid lines are given by 
		Eq.~(\ref{eq:diffusivityPorous}). }
	\label{fig:porous}
\end{figure}

As in the Laplace diffusion case, we extract the diffusivity PDF $\pi(D,t)$ and the possible values of $H$ and $\alpha$ for a given observed $\gamma=2H + \alpha - 1$ (Eq.~\ref{eq:diffusivityPorous}). Using this information, we performed numerical simulation of a   heterogeneous collection of fB walkers.
In Fig.~\ref{fig:porous}, we show the population-level picture associated to nonlinear diffusion equation for different values of $\gamma$ and the results obtained from the simulations   for corresponding values of $H$ and $\alpha$ (\ref{app:num}).

\section{Random diffusivity of  generalized distributions}
\label{sec:generalized}

In this section, we  look for the PDF of diffusivities behind generalized tracer distributions.  We focus on  two non-Gaussian diffusion processes that admit anomalous (or normal) diffusion.  The main goal  is to understand how the exponents that control the non-Gaussianity  are connected to the diffusivity of the tracers.

\subsection{Stretched diffusion}
\label{sec:stretched}

An interesting family of PDFs  is the stretched exponential distribution with temporal scaling in a $d$-dimensional space, which is written as
\begin{eqnarray}
p({\bf r},t) =\frac{1}{ \mathcal{Z}\;(D_0 t^{\gamma})^{\frac{ d}{2}} }\exp \Bigl( - \displaystyle   \Bigl[ \frac{|{\bf r}|}{ \sqrt{ D_0 t^{\gamma} } } \Bigl]^{\sigma} \Bigr),
\label{eq:stretched}
\end{eqnarray} 
in which $\sigma \in [0,2)$, $d$ is the dimension number, $\gamma$ control the anomalous diffusion, $\sigma$ control the non-Gaussianity shape, $\mathcal{Z}=\pi^{\frac{d}{2}}\Gamma(\frac{d}{\sigma}+1)/\Gamma(\frac{d}{2}+1)$ and $D_0$ is a  diffusivity-like coefficient. 
The stretched diffusion has been observed   for particles diffusing in mucin gels~\cite{cherstvy2019non}, lateral diffusion of phospholipids and proteins~\cite{jeon2016protein}, and large deviations of continuous time random walks~\cite{wang2020large}. For $\sigma=1$ and $\gamma=1$, Eq.~(\ref{eq:stretched}) becomes reduced to the $d$-dimensional Laplace distribution, that was investigated in the framework of superstatistics~\cite{postnikov2020brownian}. 
Equation~(\ref{eq:stretched}) can be written in self-similar form, by assuming that $\xi=|{\bf r}|/t^{\gamma/2}$, yielding
\begin{eqnarray}
p(\xi,t) & = & \frac{1}{ \mathcal{Z}\;(D_0 t^{2H+\alpha-1})^{\frac{
			d}{2}} } \exp \Bigl( -\Bigl[ \frac{\xi}{  \sqrt{D_0}}  \Bigl]^{\sigma} \Bigr) \label{eq:stretchedScaled}
\end{eqnarray}
where  $\gamma= 2H+\alpha-1$. Applying the Eq. (\ref{eq:stretchedScaled}) into (\ref{eq:Superdiffusivity}) with $0<\sigma<2$, we find the follow PDF for diffusivity
\begin{eqnarray}
\pi(D,t) & = & \frac{( 4\pi )^{\frac{d}{2}}  t^{(\alpha-1)(1-\frac{d}{2})} }{4 D_0^{\frac{d}{2}}\mathcal{Z} D^{2-\frac{d}{2}}}\mathcal{L}^{-1}_{s \to y} \left\{ \exp \left( - \displaystyle   \frac{s^{\frac{\sigma}{2}}}{ D_0^{\frac{\sigma}{2}} }\right) \right\}  \Bigg|_{\displaystyle y= \frac{t^{\alpha-1}}{4D} } \nonumber \\ \label{eq:Levy}
& = & \frac{\pi^{\frac{d}{2}}  t^{(\alpha-1)(1-\frac{d}{2})} }{4^{1-\frac{d}{2}} D_0^{\frac{d}{2}-1} \mathcal{Z} D^{2-\frac{d}{2}}}   L_{\sigma/2}\left( \frac{ D_0 t^{\alpha-1} }{4D}\right)  ,
\end{eqnarray}
where $L_{\sigma/2}(z)$ is the one-sided L\'evy  function. 
For $\sigma \to 1$ we have $L_{1/2}(z)= \frac{1}{2\sqrt{\pi}z^{\frac{3}{2}}}\exp\left( - \frac{1}{4z} \right)$ that implies a $\chi^2$-Gamma distribution of diffusivities 
\begin{eqnarray}
\pi(D,t)
& = &   \frac{ 4^{\frac{d}{2}} \pi^{\frac{d-1}{2}} t^{-(\alpha-1)(\frac{1+d}{2})} }{ \mathcal{Z} D_0^{\frac{d+1 }{2}} D^{\frac{1-d}{2}}} \exp\left( -t^{1-\alpha} \frac{D}{D_0} \right),  \label{eq:gammatime}
\end{eqnarray}
recalling that $\mathcal{Z} = \pi^{\frac{d}{2}}\Gamma(\frac{d}{\sigma}+1)/\Gamma(\frac{d}{2}+1)$. Special cases of Eq. ~(\ref{eq:gammatime}) that has been investigated before include: $\sigma=1$, $\alpha=1$ and $H=1/2$ for $d$-dimensional Laplace diffusion~\cite{postnikov2020brownian}; $\sigma=1$ and $d=1$ for uncorrelated noise~\cite{wang2021ergodic}. For $\sigma\neq 1$,  the  PDF (\ref{eq:Levy}) is a new result within the superstatistics of random walks.

\subsection{Power-law diffusion}
\label{sec:powerlaw}

We consider an overall  PDF of tracers that has a power-law tail, similarly to the solution of the nonlinear diffusion equation. Then, we suggest the generalized PDF 
\begin{eqnarray}
p({\bf r},t) & = & \frac{1}{\mathcal{Z}\;(\beta  t^{\gamma})^{\frac{d}{2}}}
\frac{1}{\left( 1+  \frac{{\bf r}^2}{ \beta t^{\gamma}} \right)^{\nu}}, \label{eq:nonlinearDiffusion2}
\end{eqnarray}
where $\gamma \in [0,2]$ and $\nu \in (1/d,\infty)$ are independent parameters, 
and  $\mathcal{Z}=(\pi )^{\frac{d}{2}}\Gamma \left(\left| \nu \right| -\frac{d}{2}\right)/\Gamma (\left| \nu \right| )$ allows to satisfy  $\int_0^{\infty}S_d r^{d-1}p({\bf r},t)dr = 1$ with $S_d= d \pi ^{d/2}/\Gamma \left(\frac{d}{2}+1\right)$. Here, the $\gamma$ and $\nu$  control anomalous diffusion and non-Gaussianity, respectively. 
Then, the nonlinear diffusion solution is a particular case where $\gamma$ is coupled to $\nu$. The PDF distribution of tracers (\ref{eq:nonlinearDiffusion2}) has  numerous applications in complex systems, for example in modeling of protein diffusion within bacteria~\cite{beck2021superstatistical},  parliamentary presence data~\cite{vieira2019anomalous} and  stock markets~\cite{alonso2019q}. 

To reveal the PDF of  diffusivities embodied in  power-law diffusion, we identify $\xi=|x|/t^{\gamma/2}$ in Eq.~(\ref{eq:nonlinearDiffusion2}), which  allows us to write the  self-similar form
\begin{eqnarray}
p(\xi,t) & = & \frac{1}{\mathcal{Z}\;(\beta  t^{2H+\alpha-1})^{\frac{d}{2}}}\frac{1}{\Bigl( 1+ \frac{\xi^2}{\beta}  \Bigr)^{\nu}}, \label{eq:nonlinearDiffusion22}
\end{eqnarray}
where, as before, we identified $\gamma=2H+\alpha-1$. Now, applying this self-similar solution in Eq.~(\ref{eq:Superdiffusivity}) for the $d$-dimensional case, we have that the random diffusivity obeys the  PDF
\begin{eqnarray}
\pi(D,t) & = & \frac{1}{\mathcal{Z}\;(\beta  t^{\gamma})^{\frac{d}{2}}}
\frac{( 4\pi  t^{2H} )^{\frac{d}{2}} t^{\alpha-1} }{4D^{2-\frac{d}{2}} }\mathcal{L}^{-1}_{s \to  \xi } \left\{ 1/\Bigl( 1+ \frac{s}{\beta}  \Bigr)^{\nu} \right\}  \Bigg|_{\displaystyle y= \frac{t^{\alpha-1}}{4D} },\;\;\;\
\label{eq:piinverseNon2} 
\end{eqnarray}
where $\xi=\sqrt{s}$. 
Performing the inverse Laplace transform, we have
\begin{eqnarray}
\pi(D,t) 
& = & 
\frac{(\beta t^{\alpha-1}/4)^{\nu-\frac{d}{2}}  }{ D^{\nu- \frac{d}{2}+1}\Gamma(\nu-d/2)  }\exp\Bigl(- t^{\alpha-1}\frac{\beta}{4D} \Bigr),
\label{eq:inverseGammapdf}
\end{eqnarray}
where $\nu>d/2$. The $\chi^2$-Gamma  distribution of the inverse diffusivity arises in Eq.~(\ref{eq:inverseGammapdf}), regardless of the  fractional features of the walker. Moreover,  large diffusivity fluctuations, i.e., $\pi(D) \propto D^{-\nu-1+d/2}$ with $\alpha=1$, imply a  different route to a similar asymptotic limit reported within the framework of L\'evy flights~\cite{jain2017levy}.


\section{Discussions and final remarks} 
\label{sec:final}

Random diffusivity (or mobility) has been reported in a range of experimental settings  and can be associated to different (intrinsic and extrinsic) sources. 
From the theoretical  point of view,  the impact of such variability can be  addressed through the superstatistics formalism, which brings forth interesting  insights  of how individual tracers diffusion may produce  complex population-level properties. 

In this paper,  we presented a procedure to compute the PDF of diffusivities from an  overall PDF of tracers that is self-similar. 
Our procedure considers the mixing of fractional Brownian walkers with a PDF of diffusivities that may change over time following a power-law time scaling. This approach includes different  sources of anomalous diffusion, namely, diffusivities that change over time ($\alpha \neq 1$) or 
fractional dynamics ($H \neq 1/2$), or both.  Thus, we connected a broad class of possible population density distributions to their  corresponding variability in the diffusion coefficient.  

Well-known cases, such as Laplace diffusion and nonlinear diffusion,  were investigated and used to exemplify the application of our general results. More complex cases were addressed subsequently, reaching a vast range of scenarios observed experimentally, including fat-tailed  and stretched-exponential distributions.  The generality of our results allowed us to see how certain classes of diffusion processes are interpreted in terms of diffusivity variability. For instance, we found that nonlinear-diffusion behavior is achieved when the parameters $\gamma$ and $\nu$ 
(which characterize  the scaling of the MSD and the shape of the distribution, respectively) are not independent, namely, they are related through $\frac{2\nu}{2\nu-1}=\gamma=2H+\alpha-1$, implying  that, in the distribution of the diffusivity, given by the inverse $\chi^2$-Gamma  Eq.~(\ref{eq:diffusivityPorous}), 
both the fractional exponent $H$, and also the scaling exponent $\alpha$, participate in an entangled way.  
It is worth to recall that the nonlinear diffusion is assumed to be driven by  $\partial_t p = \nabla^2 (p^{m}$), which embodies a density-dependent feedback~\cite{colombo2018},
while,
the spread of heterogeneous fBws (which is state-independent) can produce the same macroscopic features.

Despite the existing interference in the connection between individual-level mobility and the population PDF, using a proper setting and with additional  knowledge about the system, experimental approaches could benefit from our results. 
For that, it would be necessary to record the spread of a given population with sufficient temporal and spatial resolution to extract the scaling of the MSD (parameter $\gamma$) and to estimate the shape of the distribution (function $F$).   The shape should be resolved under the scaled representation, $\xi = |{\bf r}|/t^{\frac{\gamma}{2}}$, which is assumed to be preserved during the spread. Then,  performing numerically (or,
when possible, analytically) the inverse Laplace transform in Eq.~(\ref{eq:Superdiffusivity}), the distribution of diffusivities, $\pi(D)$, can be obtained. 
This is the procedure we followed, from a theoretical perspective, for paradigmatic diffusion pictures (Figs.~\ref{fig:laplace} and \ref{fig:porous}). 

We highlight the fact that the overall MSD of tracers carries  ambiguity about the source of anomalous diffusion, i.e.,  $  \langle {\bf r}^2 \rangle \propto t^{\gamma}$  with $\gamma=2H+\alpha-1$,  being indistinguishable which interpretation leads to the  anomalous diffusion exponent. To know precisely what is the anomalous diffusion mechanism,  complementary information about tracer's trajectories (Fig.~\ref{fig:trajectories}) and environment spatiotemporal structure need to be provided.

Future research should focus on keep extending the connection between diffusivity variability and population distribution. 
First, if the population PDF, $p({\bf r},t)$, were  separable as $p({\bf r},t) =  F(\xi)/[g(t)]^{d} $, with $\xi= |{\bf r}|/g(t)$,   where  $g(t)$ can be an arbitrary function (not necessarily a power-law in time), then, our approach based on the Laplace transform formalism could still be used, 
extending the scope of applicability beyond the addressed self-similar scenario.
In the current case, we have considered $g(t) \propto 
\lambda(t) t^{2H}$ with $\lambda(t) \propto t^{\alpha-1}$. But, 
if the walkers were not of a fB type, or if diffusion time-dependency $\lambda(t)$ were not a power-law, then, the ambiguity about their contributions in spreading might be broken. 
As a consequence, one could infer more information about the microscopic dynamics. 
This discussion suggests natural extensions worth of future investigation, for instance, the one in which $\lambda(t) \propto e^{\pm\lambda_0 t}$~\cite{cherstvy2021anomalous}.

Another step forward would be to address the mixing of diffusion process with  short  or long tails, such as  $q$-Gaussian~\cite{tsallis2009introduction} with $q<1$ (finite support) and $q>1$ (power-law tails), respectively, that by itself introduce a connection between the shape of the population distribution and the scaling of the MSD. 
Finally, it would be also interesting to explore other forms of   heterogeneity, e.g., in the $\alpha$ exponent of the scaling $\lambda(t)\propto t^{\alpha-1}$  or in the Hurst exponent $H$ \cite{beck2021superstatistical}, which in our case are the same for all the walkers.

{\bf Acknowledgments:}  
CA acknowledges Brazilian agency CNPq (process 311435/2020-3) for partial financial support. MAFS and CA acknowledge CAPES (finance code 001).


\appendix

\section{Microscopic dynamics}
\label{app:num}

In  individual-level simulations, we generate the trajectories of
each fB walker $i=1,2,\ldots,N$ from microscopic rules. The fractional component is related to the Hurst exponent $H$, and each individual is associated to a different diffusion coefficient, $D_i(t) = \mu_i \lambda(t)$,  where $\mu_i$  is a mobility constant in time but different for each walker, and $\lambda$ is the deterministic protocol, common to all walkers. The heterogeneity of the ensemble is given by sampling the values of $\mu_i$ from a probability distribution $\theta(\mu)$ at the beginning of our simulations.

Then, the microscopic dynamics of each walker is evolved through a Langevin-like equation that incorporates fractional motion and time scaling, as
\begin{eqnarray}
\dot{x} = \sqrt{2\mathcal{D}(t)}\mathbf{\eta}_{H}(t),
\label{eq:langevinWalker}
\end{eqnarray}
where  $\mathcal{D}(t)= \mu \alpha t^{\alpha-1}/4$~\cite{wang2021ergodic} and $\eta_{H}$ is the fractional noise with Hurst exponent $H$~ \cite{mandelbrot1968fractional}, 
generated by using the so-called Davies–Harte  algorithm~\cite{davies1987tests}.  


%

\section{ Anomalous diffusion}
\label{app:anomalous}

The general $\mu$-moment, according to  the superstatistical mixture of fB walkers defined in Eq.~(\ref{eq:superstatisticalScaled}),   is given by 
\begin{eqnarray}
\langle |{\bf r}|^{\mu} \rangle  
& = &   \int_0^{\infty} \int_{|{\bf r}|} |{\bf r}|^{\mu} \frac{4}{\lambda(t)}\pi_s\left(\frac{\lambda}{D} \right) \mathcal{G}_H({\bf r},t|D) |{\bf r}|^{d-1} S_d d{\bf r} dD 
\end{eqnarray} 
where $S_d= d \pi ^{d/2}/\Gamma \left(\frac{d}{2}+1\right)$. 
Performing the following calculations
\begin{eqnarray}
\langle |{\bf r}|^{\mu} \rangle  &=&  \int_0^{\infty} \int_{|{\bf r}|} \frac{|{\bf r}|^{\mu}}{(4 t^{2H} \pi D)^{\frac{d}{2}}} \exp\left( - \frac{{\bf r}^2}{4 t^{2H} D} \right) \frac{\pi_s(\lambda(t)/D)}{\lambda(t)}|{\bf r}|^{d-1} S_d d{\bf r} dD \nonumber \\
& = &  \int_0^{\infty}\frac{\pi_s(\lambda(t)/D)}{\lambda(t)} \int_{|{\bf r}|} \frac{|{\bf r}|^{\mu}}{(4 t^{2H} \pi D)^{\frac{d}{2}}} \exp\left( - \frac{{\bf r}^2}{4 t^{2H} D} \right) |{\bf r}|^{d-1} S_d d{\bf r} dD \nonumber \\
& = & t^{H\mu }\int_0^{\infty}\frac{\pi_s(\lambda(t)/D)}{\lambda(t)} \frac{d 2^{\mu -1} D^{\frac{\mu }{2}} \Gamma \left(\frac{d+\mu }{2}\right)}{\Gamma \left(\frac{d}{2}+1\right)} dD  \label{eq:rmu} \end{eqnarray} 
and redefining a new integration variable $\overline{r}=|{\bf r}/t^H|$, Eq.~(\ref{eq:rmu})  implies
\begin{eqnarray}
\langle |{\bf r}|^{\mu} \rangle & = &  t^{H \mu} (\lambda(t))^{\frac{\mu}{2}} \frac{ 2^{\mu -1} d D^{\frac{\mu }{2}} \Gamma \left(\frac{d+\mu }{2}\right)}{\Gamma \left(\frac{d}{2}+1\right)} \int_0^{\infty}\pi_s(y) y^{-2-\frac{\mu}{2}} dy.
\end{eqnarray} 
Assuming the power-law temporal behavior for  $\lambda(t)\sim t^{\alpha-1}$, as argued in the main text, we obtain
\begin{eqnarray}
\langle |{\bf r}|^{\mu} \rangle  & \sim &     t^{(2H+\alpha-1)\frac{\mu}{2}}\,,
\label{eq:nmoment}
\end{eqnarray}
as presented in Sec~\ref{sec:approach}.

\section{ PDF of diffusivities for nonlinear diffusion}
\label{app:porous}

Starting from Eq.~(\ref{eq:piinverseNon}), 
presented in Sec.~\ref{sec:porous}, our goal is to use the equation
\begin{eqnarray}
\pi(D,t) = \frac{( 4\pi t^{\alpha-1} )^{\frac{1}{2}}  }{4 D^{\frac{3}{2}}\gamma_{\nu}\beta_{\nu}}\mathcal{L}^{-1} \left\{ \left( 1+ \frac{\xi}{\nu\beta^2_{\nu}}  \right)^{-\nu} \right\}  \Bigg|_{\displaystyle y= \frac{t^{\alpha-1}}{4D} } 
\end{eqnarray}
to find Eq.~(\ref{eq:diffusivityPorous}).  Using the Laplace inverse transform, we have
\begin{eqnarray}
\pi(D,t) & = &  \frac{( 4\pi t^{\alpha-1} )^{\frac{1}{2}}  }{4 D^{\frac{3}{2}}\gamma_{\nu}\beta_{\nu}}  \left[ y^{\nu-1} (\nu \beta_{\nu}^2 )^{\nu} \frac{\exp\left(- y \nu \beta_{\nu}^2 \right)}{\Gamma(\nu)} \right]  \Bigg|_{\displaystyle y= \frac{t^{\alpha-1}}{4D} } \nonumber \\
& = &   \frac{\nu^{\nu}\sqrt{4\pi t^{\alpha-1} }}{4\sqrt{ D^{3}} \gamma_{\nu}\beta_{\nu}^{1-2\nu} }  \left[ y^{\nu-1} \frac{\exp\left(- y \nu \beta_{\nu}^2 \right)}{\Gamma(\nu)} \right]  \Bigg|_{\displaystyle y= \frac{t^{\alpha-1}}{4D} } 
\end{eqnarray}
that implies
\begin{eqnarray}
\pi(D,t) & = &  \frac{(\nu \beta_{\nu}^2/4)^{\nu-\frac{1}{2}} }{D^{(\nu-1/2)+1} } \frac{t^{(\alpha-1)(\nu-1/2)}}{\Gamma(\nu-\frac{1}{2})} \exp\left(-  \frac{\nu \beta_{\nu}^2 t^{(\alpha-1)}}{4D} \right).
\end{eqnarray}
Defining a new parameter  $\overline{\nu}=\nu-1/2$, we write the inverse $\chi^2$-Gamma distribution with time dependence
\begin{eqnarray}
\pi(D,t) & = &  \frac{1}{ \Gamma(\overline{\nu}) } \frac{\left(\nu t^{(\alpha-1)}\beta_{\nu}^2/4  \right)^{\overline{\nu}} }{D^{1+\overline{\nu}}} \exp\left(-\frac{\nu t^{(\alpha-1)} \beta_{\nu}^2 }{4D}\right) , 
\end{eqnarray}
as shown in Eq.~(\ref{eq:diffusivityPorous}).

\end{document}